\documentclass[runningheads]{llncs}
\usepackage{graphicx}
\usepackage{cite}
\usepackage{url}
\usepackage{amsmath}
\usepackage{tabularx}

\usepackage{array}
\usepackage{xspace}
\usepackage{blkarray}
\usepackage{multirow}
\usepackage{subfig}
\usepackage{graphicx}
\usepackage{epstopdf}
\usepackage{wrapfig}

\usepackage[table,xcdraw]{xcolor}

\usepackage{textcomp}
\usepackage{gensymb}
\usepackage{siunitx}
\usepackage{rotating}
\usepackage{multirow}
\usepackage{commath}
\usepackage{multicol}
\usepackage{multirow}
\usepackage{blkarray}
\usepackage{relsize}
\usepackage{booktabs}
\usepackage{slashbox}
\usepackage{multirow}

\usepackage{soul}

\usepackage{amssymb}
\usepackage{pifont}
\usepackage{color}

\newcommand{\sysname}{\textsc{IoTguard}\xspace}

\newcommand{\classM}{\textit{malicious}\xspace}
\newcommand{\classB}{\textit{benign}\xspace}

\newcommand{\wifi}{WiFi\xspace}

\newcommand{\psweep}{\mbox{P-Sweep}\xspace}
\newcommand{\pscan}{\mbox{P-Scan}\xspace}

\begin{document}
\title{Real-time IoT Device Activity Detection in Edge Networks}

\author{Ibbad Hafeez\inst{1} \and
Aaron Yi Ding\inst{2,3} \and
Markku Antikainen\inst{1,4} \and \\
Sasu Tarkoma\inst{1,4}}
\authorrunning{I. Hafeez et al.}
%
\institute{
University of Helsinki, Helsinki, Finland,
\email{firstname.lastname@helsinki.fi}
\and
Technical University of Munich, Munich, Germany,
\email{aaron.ding@tum.de}\\
\and 
Delft University of Technology, Delft, Netherlands,\\
\and
Helsinki Institute of Information Technology, Helsinki, Finland}
\maketitle              
\begin{abstract}
The growing popularity of Internet-of-Things (IoT) has created the need for network-based traffic anomaly detection systems that could identify misbehaving devices.
In this work, we propose a lightweight technique, \sysname, for identifying malicious traffic flows.  
\sysname uses semi-supervised learning to distinguish between \classM and \classB device behaviours using the network traffic generated by devices. 
In order to achieve this, we extracted 39 features from network logs and discard any features containing redundant information.
After feature selection, fuzzy C-Mean (FCM) algorithm was trained to obtain clusters discriminating \classB traffic from \classM traffic. 
We studied the feature scores in these clusters and use this information to predict the type of new traffic flows. 
\sysname was evaluated using a real-world testbed with more than 30 devices.
The results show that \sysname achieves high accuracy ($\ge98\%$), in differentiating various types of \classM and \classB traffic, with low false positive rates. 
Furthermore, it has low resource footprint and can operate on OpenWRT enabled access points and COTS computing boards. 

\end{abstract}

\section{Introduction}
\label{sec:introduction}

The Internet-of-Things (IoT) trend has significantly increased the number devices connected to the Internet. Predictions forecast this number to exceed 20 billion by year 2020~\cite{IoTdeviceForecast1}. Despite its benefits, a number of security concerns have been raised about the connected devices themselves. 
Majority of smart devices operate on limited power and computational resources and hence do not support host-based security software such as anti-malware. Also, IoT products are mostly developed by product development teams who have limited resources and who may not follow standard security practices e.g. reusing code snippets, weak encryption keys, lack of security-by-design etc.~\cite{Senrio:SingleFlaw, DLink:SingleFlaw, 6975580}.

IoT devices are lucrative targets for attackers who want to obtain user-related information or to perform large scale network attacks. Due to the poor security of many IoT devices, network-based security solutions are often the only line of defence against incoming attacks that target these devices. 

Unfortunately, traditional network security solutions, such as network intrusion detection/prevention systems (NIDS/NIPS) and firewalls, fall short in distinguishing and filtering malicious traffic generated by these smart devices for a number of reasons.  Firstly, it is infeasible to collect signatures for all possible network interactions for these devices, due to heterogeneity in devices and firmware versions. In practice, a device's network behaviour may vary significantly in different firmware releases. Secondly, the costs of deploying and maintaining traditional NIDS/NIPS and firewall solution is high for small-office and home networks. Lastly, amount of network traffic data that needs to be processed may overwhelm the NIDS/NIPS systems that perform traffic analysis.
Therefore, it is necessary to research new solutions for traffic monitoring and classification, which are self-adaptive, cost efficient and do not require specialized hardware. 

In this work, we propose a self-adaptive semi-supervised learning based classification scheme named as \sysname, which predicts traffic class (i.e. \classM or \classB) based on the network activity of the device generating the traffic. 

Our technique primarily uses the data extracted from network logs that are obtained from access-points (APs) and gateways. \sysname does not specifically rely on specialized logs obtained from domain-controllers, firewalls or NIDS, because smaller networks (i.e. small-office and home networks, aka.\ SOHO networks) rarely have these. However, if available, our technique can use data also from such specialized logs to further improve the efficiency and accuracy of the system. 
All this data is combined to identify network-level patterns for different kinds of traffic the devices generate, and use these patterns to identify any malicious activities in the network. We resolved data imbalance issues by oversampling and under-sampling data from minority and majority class respectively.
Our choice of unsupervised learning is motivated by the reason that class labels for most network logs are not available and classification scheme should be able to learn from various patterns observed in network traffic. 

Our work demonstrates that a simple, yet effective, clustering technique combined with in-depth feature analysis enables real-time traffic classification, without requiring dedicated hardware. 
Our key contributions are:
\begin{itemize}
\item We propose a pipeline detailing feature extraction, analysis and reduction techniques, to develop the set of most useful features for performing clustering on network data.

\item We propose traffic classification scheme using fuzzy C-Mean clustering and fuzzy interpolation scheme, which is able to determine the degree of maliciousness, therefore, giving more information for taking appropriate measures to handle different types of malicious traffic.

\item We evaluate the performance of \sysname in real-world environment with off-the-shelf consumer-grade devices. \sysname boasts high prediction accuracy ($\ge98\%$) for both \textit{binary} and \textit{multi-class} problems with low false-positive-rate ( $\simeq 0.01$).
\end{itemize}

In the rest of paper, Sect.~\ref{sec:threat_model} discuss our threat model outlining the types of attacks in IoT edge networks, followed by methodology in Sect.~\ref{sec:method}. The data set and evaluation results are discussed in Sect.~\ref{sec:dataset} and Sect.~\ref{sec:evaluation} respectively. Section~\ref{sec:stoa} gives a comparison of \sysname with existing approaches. Section~\ref{sec:discussion} discusses the some limitations of \sysname, followed by concluding remarks.

\section{Threat model}
\label{sec:threat_model}
This work focuses on small-office and home networks. These networks usually have a star topology where all devices are connected to an access point that also provides Internet connectivity. IoT devices in such networks are often mismanaged and not hardened. Thus, while the devices are intrinsically benign, an attacker can easily compromise and use these devices for various follow-up attacks.
The list of attacks, which can be launched in a SOHO networks with an aid of an already compromised device, is given as:

\textbf{\textbf{Network-scanning attacks}}, where an adversary tries to find any device on the network, running services with open, unguarded ports. Network scanning commonly include \textit{port-scan}, \textit{port-sweep} and \textit{address-sweep} attacks.

\textbf{\textit{Flooding attacks}}, where a (compromised) device participates in a large scale Distributed Denial-of-Services (DDoS) attack. DDoS attacks are often used as a smoke-screen to divert attention off dedicated attacks occurring in parallel.

\textbf{\textit{Infection attacks}}, where a compromised or infected device actively tries to infect to other devices in the network. For example, an attacker may try to make repeated login attempts to services discovered by \textit{network-scanning}, in order to download malware on other devices in the network.

\textbf{\textit{Spying attacks}}, where a device collects user data without explicit consent and sends it to untrusted third party. 

This work uses network-level semantics of these attacks to predict type of traffic in the network. It does not individually profile each device's behaviour as \classB or \classM. Instead, it uses feature scores observed in various traffic types, to identify traffic, irrespective of what device generated it. 

\section{Methodology}
\label{sec:method}

\subsection{Design challenge}
\label{sec:overview}
A key limitation in using supervised learning algorithms for network traffic classification is the unavailability of labelled data covering all traffic classes seen in real-world environments. 
With a huge variety of IoT devices and their heterogeneous mode of operations, it is expensive and infeasible to label all data collected by monitoring network traffic. 
To overcome this challenge, unsupervised learning provides a better alternative, as it does not require nor depend on labelled data. Clustering can partition large volumes of network traffic data into small number of clusters based on similar patterns observed in data.

Any new data point will be added to a cluster based on its similarity with existing data points in the cluster. Meanwhile, these clusters can be rearranged, divided or combined depending on number of classes of data.

\subsection{Feature extraction}
\label{sec:feature_extraction}

\sysname uses features collected from the access point and, if available, from individual device logs. With an assumption of unique IP address per device, we use the source and destination IP addresses together with timestamps (3-tuple identifier) from each traffic flow, for identifying the feature vector for that flow.
Each feature vector consists of a total of 39 discrete and continuous features, listed in Table~\ref{tbl:features}.

\begin{table}[]
\caption{Discrete and continuous features extracted from network connections} 
\centering
\label{tbl:features}
\begin{tabular}{p{0.04\linewidth}|p{0.16\linewidth}p{0.80\linewidth}}
\hline
\multicolumn{1}{c}{} & \multicolumn{1}{c}{\textbf{Type}}  & \multicolumn{1}{c}{\textbf{Feature}} \\\hline

\multirow{5}{*}{\rotatebox[origin=c]{90}{\textbf{Discrete}}} & L2 Protocol & ARP, LLC \\
& L3 Protocol &  IP, ICMP, ICMPv6, EAPoL  \\ 
& L4 Protocol & TCP, UDP  \\
& L5 protocol & HTTP, HTTPS, DHCP, BOOTP, SSDP, (M)DNS, NTP \\ 
& IP Options  & Padding, Router Alert \\ \hline

\multirow{10}{*}{\rotatebox[origin=c]{90}{\textbf{Continuous}}}  &
   Src and dest 	& \# unique destination IP addresses \\ 
& 				& \# unique source and destination ports  \\ \cline{2-3}
& Counters 		& \# total connections, \# connections to/from unique dest/src \\
&				& Connection lengths,\ SYN packets \& errors, REJ errors, URG packets \\ \cline{2-3}
& Data  			& Total data transferred. \\
&				& Total data from source to destination \\ 
&				& Total data from destination to source \\
&				& Packet sizes, payload signatures  \\ \cline{2-3}
& Auth.			& Total login attempts (inc. SSH connection, using default credentials, failed login attempts)                                                                                               \\ \bottomrule
\end{tabular}
\end{table}

For some features (e.g. authentication and network discovery), we need accurate time synchronization among all devices. In case if network does not use time synchronization mechanisms such as NTP, we have to manually account for the time differences between network and device logs. 

We aggregate the same-host, same-service features over $n$ latest connections instead of using time-based aggregation. Time-based aggregation (used in KDDCup99 dataset~\cite{KDDCup99Dataset}) aggregates the features over a definite time e.g. number of connections made in last two seconds between \textit{Device-A} and \textit{Device-B}. 
This scheme falls short in detecting attacks where attacker introduces a time-delay between successive connection attempts. In contrast, connection-based aggregation techniques aggregate features over last $n$ connections i.e. out of last $n$ connections made by \textit{Device-A}, how many terminated at \textit{Device-B}. This technique accommodates the time-delay added to successive connections. However, if $n$ is small and device connects to several destinations simultaneously i.e. behaviour not observed commonly in compromised devices targeting certain destination, connection-based aggregation may not work effectively.

\subsection{Feature analysis}
\label{sec:feature_analysis}

The value distributions of the features was studied in order to identify relative importance of features, based on variance and modality. Any features with low variance across different samples are discarded because they do not substantially contribute to clustering. This dimensionality reduction also helps speed up the clustering process.

\begin{figure}
\begin{minipage}{.33\linewidth}
\centering
\subfloat{\label{fig:cdf_dest_ip}\includegraphics[scale=.21]{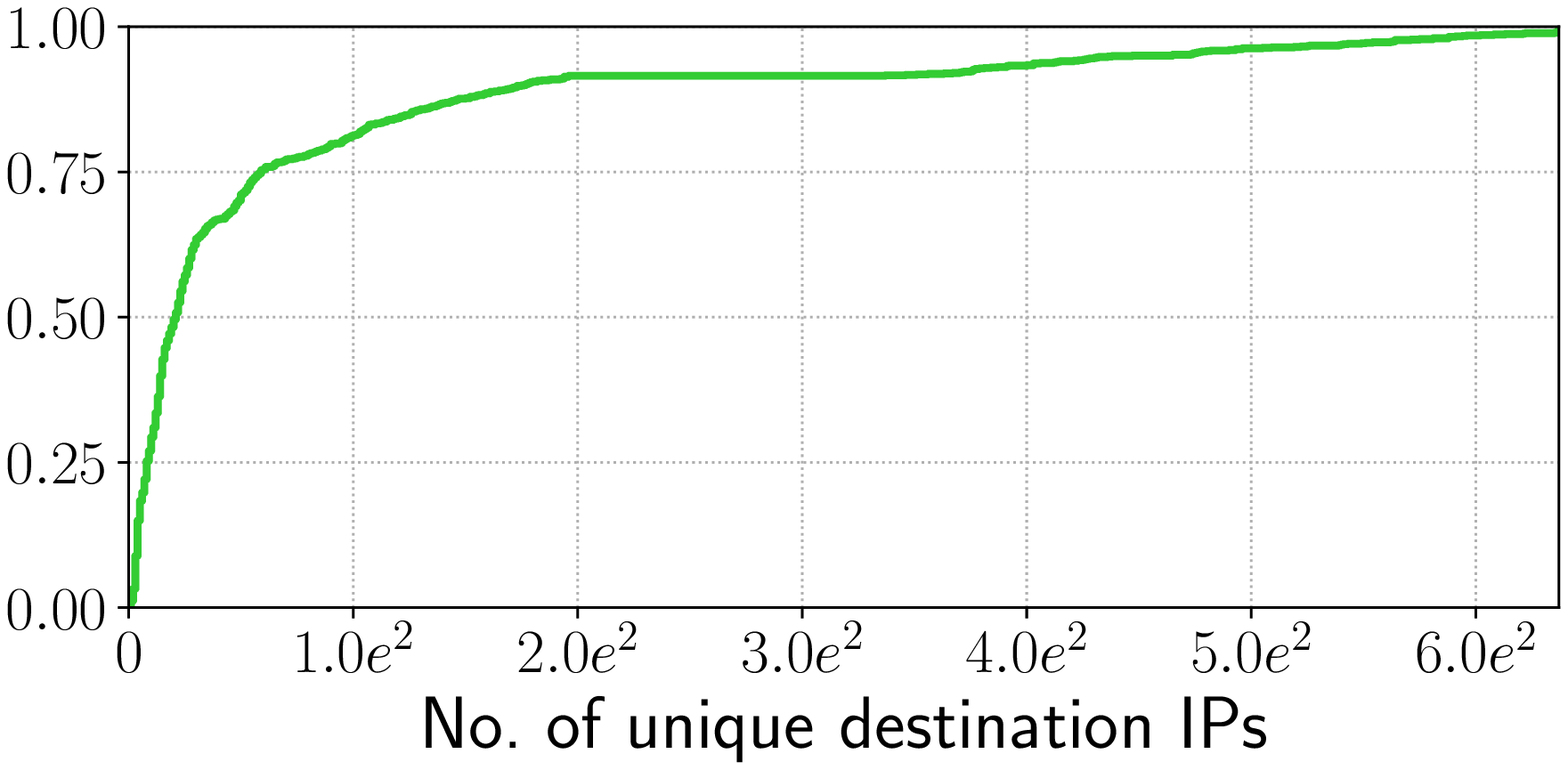}}
\end{minipage}%
\begin{minipage}{.33\linewidth}
\centering
\subfloat{\label{fig:cdf_dest_port}\includegraphics[scale=.21]{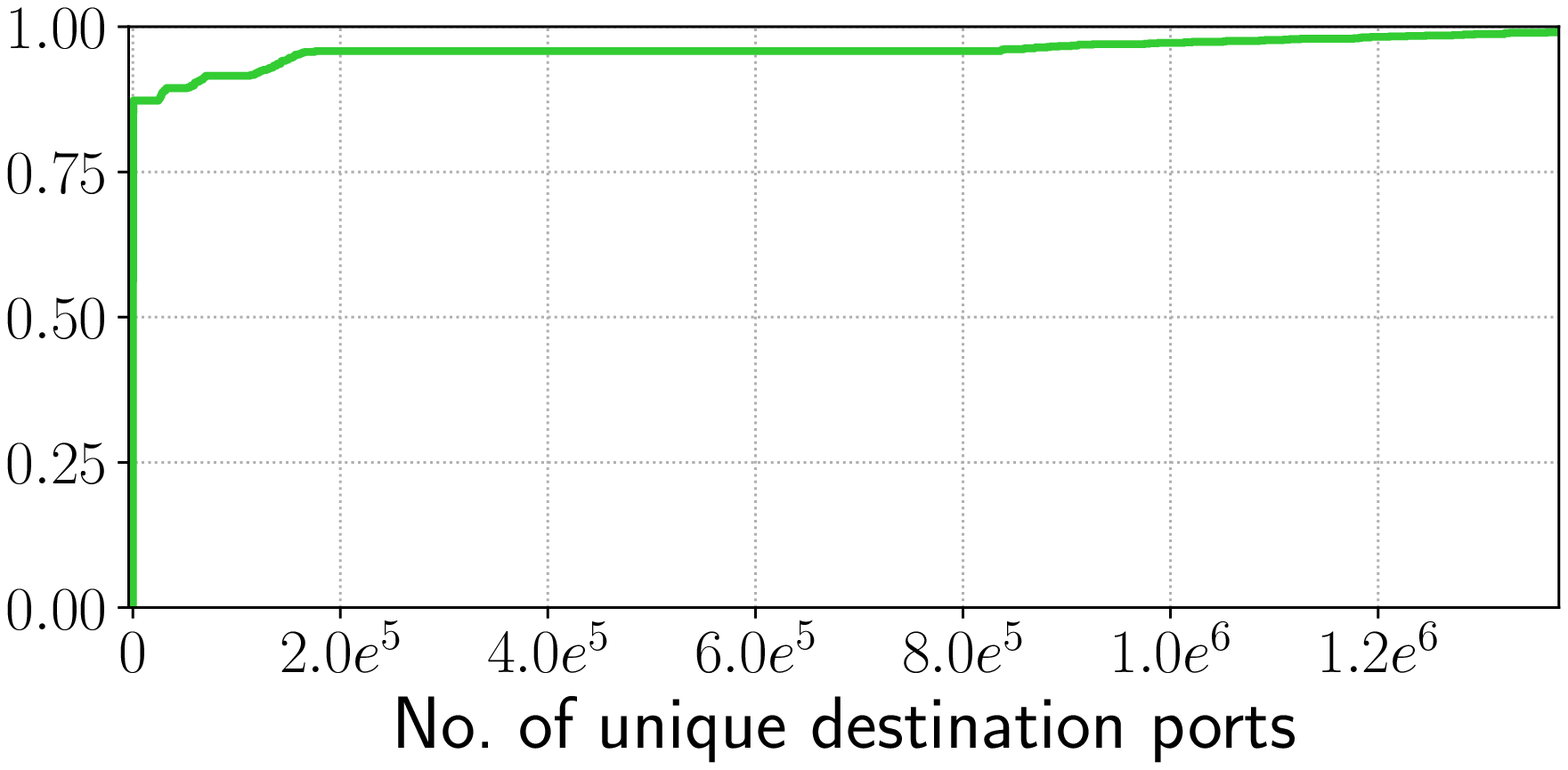}}
\end{minipage}%
\begin{minipage}{.33\linewidth}
\centering
\subfloat{\label{fig:cdf_conn}\includegraphics[scale=.21]{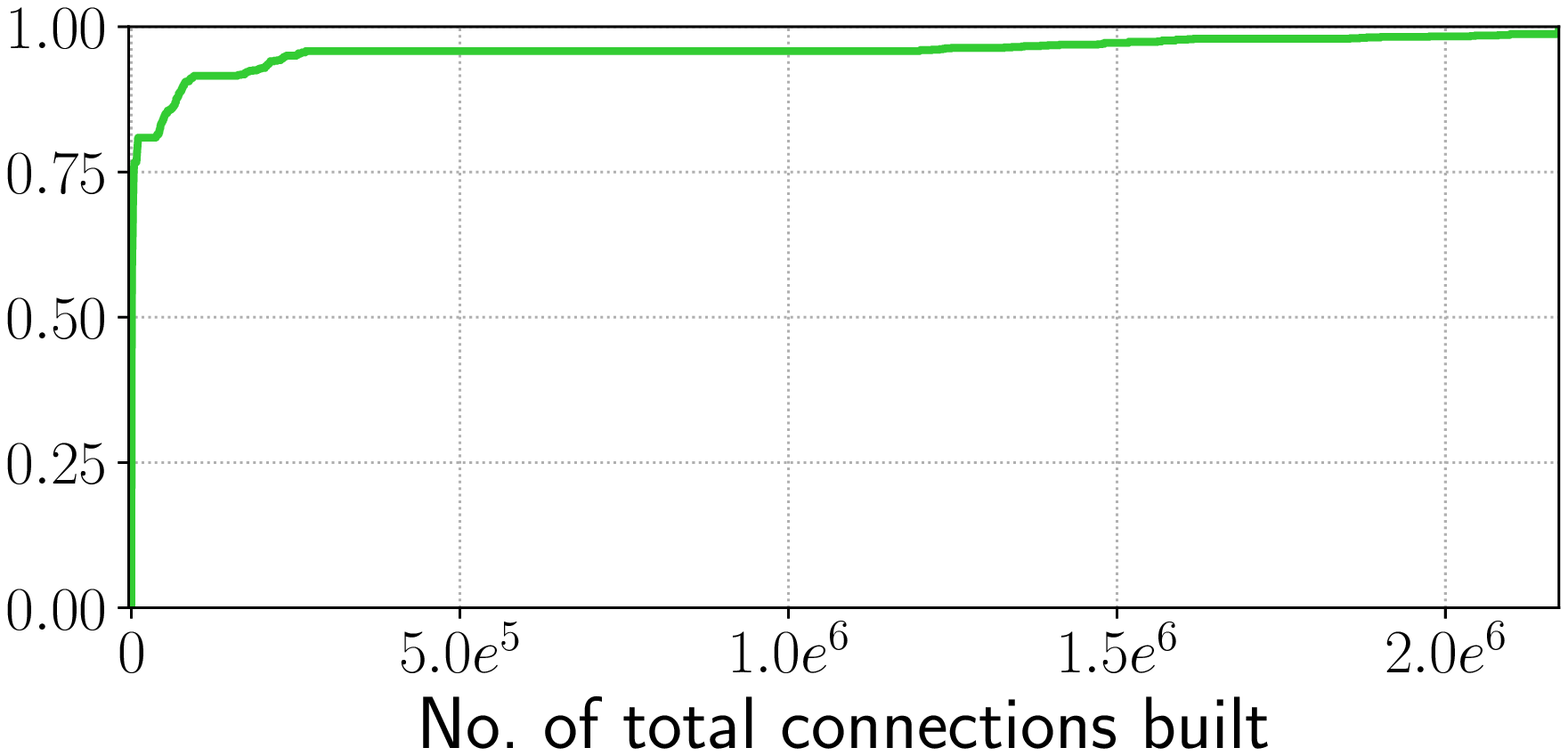}}
\end{minipage}
\caption{CDF plots for a subset of connection metadata}
\label{fig:cdf_features}
\end{figure}

Figure~\ref{fig:cdf_features} shows cumulative distribution functions for three (of 39) extracted features. The distributions in the figures are not Gaussian, but heavy tailed with majority of probability mass lying in smaller values. For example Figure~\ref{fig:cdf_dest_ip} shows that $\ge70\%$ devices connect to $\le20$ unique destinations but there are some devices which connect to $\ge6000$ unique destinations. 
The tail of these distributions is particularly interesting because it encapsulates events where a device may be exhibiting anomalous behaviour. The knowledge from feature value distributions is used to choose the features that will most likely result in clusters with well defined boundaries and outliers. 

\subsection{Feature reduction}
\label{sec:feature_reduction}
Feature reduction decreases the model complexity, reduces resource consumption, and improves generalization. 
Therefore, \textit{correlation-based feature selection} (CFS), \textit{deviation method}, and feature value distributions are used to identify and remove any features that do not significantly contribute to clustering process. 

Pearson coefficient provides fairly accurate results with bounded feature value ranges when size of the dataset is large~\cite{Bohara:2016:IDE:2898375.2898400}. We use Pearson correlation coefficient $R$ to measure the linear dependencies of strongly correlated features.
One of any two strongly correlated features (i.e.\ $R \ge 0.99$) can be discarded as redundant.


With deviation method, we first mine 1-length items from each feature to obtain 39 feature vectors that contain frequent items for each of the seven activity types listed in Table~\ref{tbl:scenarios}. The frequent items for binary features can be found with algorithms such as Apriori~\cite{Agrawal:1994:FAM:645920.672836} or FP-Growth~\cite{NARVEKAR2015101}. For continuous variables, 1-length items are found by comparing the frequency of a continuous variable against a specified minimum support. If the deviation range for a feature overlaps across all traffic types, we do not expect it to significantly contribute in clustering and, therefore, remove it. 

The normalized feature scores is studied for every feature in all clusters and any features with similar values across differnet clusters are removed. To ensure that no feature over-influences clustering, all feature values are normalized to range $\lbrack0,1\rbrack$.
It was observed that the use of \textit{principal component analysis} (PCA) for dimensionality reduction prior to clustering does not benefit to our approach because PCA fails to capture outliers (tail of distribution) in its principal components and those outliers can be particularly useful for identifying anomalies.

\subsection{Clustering}
\label{sec:clustering}

Fuzzy C-mean (FCM) clustering algorithm is used to separate data points based on their self similarity. 
Our choice of FCM is based on its ability to maintain weighted association of any point not only for the cluster which it is assigned to, but for neighbouring clusters as well, where it is weakly associated~\cite{5370013}. These weak associations are useful in predicting labels for unknown traffic flows since \textit{all} cluster  associations are considered when assigning assigning a label. This approach helps in reducing the number of false-positives.

Using FCM, initially a random membership value is assigned to each data point $X_{j}\;(j=1, 2, ..., n)$ for every cluster $C_{i}\;(i=1,2,...,c)$. Each data point $X_{j}$ is represented as $\left(f_{j}^{(1)}, f_{j}^{(2)},..., f_{j}^{(k)}, ..., f_{j}^{(h)}\right)$ where $f_{j}^{(k)}$ is value for $k^{th}$ feature in $X_{j}$ and $1\leq k \leq 39$ (i.e.~39 features).

The membership value $\mu_{ij}$, ($0 \leq \mu_{ij} \leq 1$) for a data point $X_{j}$ assigned to cluster $C_{i}$ is such that $\sum_{i=1}^{c} \mu_{ij}=1$ for ${1\le i \le c}$ and $1\le j \le n$. The membership values $\mu_{ij}$ and cluster centres $V_{i}$ are optimized using Eq.~\ref{eq:cluster-centers}, to minimize objective function in Eq.~\ref{eq:obj-ftn}.

\begin{align}
\mathsmaller{
\mu_{ij} =
\left(
\sum\limits_{d=1}^{c}
     \left(\dfrac{\norm{V_{i}-X_{j}}}{\norm{V_{d}-X_{j}}}\right)^{\dfrac{\mathsmaller{2}}{\mathsmaller{m-1}}}
\right)^{-1}
}
;\hspace{2mm} &
\mathsmaller{
	V_{i} = \dfrac{\sum\limits_{j=1}^{n}(\mu_{ij})^{m}\times X_{j}}{\sum\limits_{j=1}^{n}(\mu_{ij})^{m}}
}
\hspace{2mm};\hspace{2mm} &  
\mathsmaller{\genfrac{}{}{0pt}{}{1 \leq i \leq c}{1 \leq j \leq n}}
\label{eq:cluster-centers}
\end{align}

\begin{equation}
J_{m}=\sum_{i=1}^{c}\sum_{j=1}^{n}\mathlarger{\mu}_{ij}^{m}\norm{V_{i}-X_{j}}^2
\label{eq:obj-ftn}
\end{equation}
where $m$ is fuzziness index~\cite{Zhou2014} and $\norm{V_{i}-X_{j}}$ is the Euclidean distance between cluster center $V_{i}$ for cluster $C_{i}$ and data point $X_{j}$.

Clusters labels are assigned based on feature value distribution for each cluster. The labels can be manually verified using dataset ground truth. Each of these clusters is translated to a fuzzy rule, used by \textit{fuzzy interpolation scheme} (FIS) for predicting type of given traffic flow. 

\subsection{Parameter optimization}
\label{sec:param_selection}
The optimal number of clusters $i$ is determined by examining the degree of cohesion among data points in a cluster, \textit{fuzzy partition coefficient}~\cite{TRAUWAERT1988217} (FPC), and trade-off between sensitivity, specificity and accuracy of our prediction. 

The process is initialized with a range of possible values for $i$. Then, FCM algorithm runs for $n=3000$ iterations to calculate FPC and \textit{within-cluster-sums-of-distances} (WCSD) for each value of $i$. 
$i$ with minimum WCSD is chosen, to remove any initialization bias and prevent the output to reside in local minima~\cite{Bohara:2016:IDE:2898375.2898400}. WCSD is calculated using Eq.~\ref{eq:wcsd}, where $c$ is the number of clusters, $S_{i}$ is the set of data points belonging to $i^{th}$ cluster, and $x_{ki}$ is the $k^{th}$ variable of $V_{i}$. 

\begin{equation}
WCSD = \sum\limits_{i=1}^{c}\sum\limits_{j \in S_{i}}\sum\limits_{k=1}^{p} \norm{x_{ki} - x_{ji}}
\label{eq:wcsd}
\end{equation}  

Silhouette values~\cite{4426662} (using Eq.~\ref{eq:silhouette_value}) are calculated for all data points $x_{k}$ and verify our choice of $i$ by studying how well a given data point belongs to the cluster it is assigned to. The optimal choice for $i$ will have minimum WCSD and maximum average silhouette value. 
\begin{equation}
s(x) = \frac{b(x)- a(x)}{\max(a(x),b(x))}
\label{eq:silhouette_value}
\end{equation}

\subsection{Prediction algorithm}
\label{sec:prediction}

\sysname uses \textit{fuzzy interpolation scheme} (FIS) to predict the type of traffic using the rules obtained from clustering. FIS allows us to deduce a conclusion using a sparse fuzzy rule base. 
Let us consider an sparse fuzzy rule set such as

$\mathsmaller{Rule\;1:\; if\;f_{1} \in A_{11}, f_{2} \in A_{21}, \; ... \;, f_{k} \in A_{k1}, \; ... \;, f_{h} \in A_{h1}\;\Longrightarrow\;y\;\in\;O_{1}}$

$\mathsmaller{Rule\;2:\; if\;f_{1} \in A_{12}, f_{2} \in A_{22}, \; ... \;, f_{k} \in A_{k2}, \; ... \;, f_{h} \in A_{h2}\;\Longrightarrow\;y\;\in\;O_{2}}$

\vspace{-0.5mm}
\hspace{0.20\textwidth}\vdots

\vspace{-1.5mm}
$\mathsmaller{Rule\;Q:\; if\;f_{1} \in A_{1q}, f_{2} \in A_{2q}, \; ... \;, f_{k} \in A_{kq}, \; ... \;, f_{h} \in A_{hq}\;\Longrightarrow\;y\;\in\;O_{q}}$

$\mathsmaller{Observation:\;f_{1} \in A_{1}^{*},\; f_{2}\in A_{2}^{*}, \; ... \;, f_{k}\in A_{k}^{*}, \; ... \;, f_{h} \in A_{h}^{*}}$

\vspace{-2mm}\rule{8cm}{0.4pt}

\vspace{-0.5mm}
$\mathsmaller{Conclusion:\;y = O^{*}}$

\noindent
where $R_{i}$ ($1 \leq i \leq Q$) is $i^{th}$ rule in sparse fuzzy rule base generated from cluster $C_{i}$. 

$A_{ki}$ and $O_{i}$ are triangular fuzzy sets for $k^{th}$ antecedent feature $f_{k}, 1 \leq k \leq h$ and consequent variable $y$ respectively. For any new observation, $A_{k}^{*}$ and $O^{*}$ are triangular fuzzy sets for antecedent and consequent variable obtained as a result of interpolation of spare fuzzy rule base. 

The classification rules obtained from clusters generate the rule base such that $R_{i}$ is generated from $C_{i}$ with $h$ antecedent features and one consequent label assigned to the given cluster. 

\begin{center}
$\mathsmaller{R_{i}:\; \mathit{if}\;f_{1} \in A_{1i}, f_{2} \in A_{2i}, \; ... \;, f_{k} \in A_{ki}, \; ... \;, f_{h} \in A_{hi}\;\Longrightarrow\;y\;\in\;B_{i}}$
\end{center}

The characteristic points $a_{ki}$, $b_{ki}$, $c_{ki}$ for triangular fuzzy set are calculated for all antecedents $A_{ki}$ and consequent $B_{i}$ in $R_{i}$.
The weight $W_{i}$ of given rule $R_{i}$ $\left(i=1, 2, ..., c\right)$ is calculated on the basis of input observations $x_{1}=f_{j}^{(1)}, x_{2}=f_{j}^{2)}, ..., x_{h}=f_{j}^{(h)}$ as: 

\begin{equation}
\mathsmaller{
    W_{i} = \left(\mathlarger{\sum}\limits_{d=1}^{c} \left(\dfrac{\norm{r^{*} - r_{i}}}{\norm{r^{*} - r_{d}}} \right)^{2} \right)^{-1}
},
\label{eq:calc_weight}
\end{equation}
where $r^{*}$ is the input feature vector $\left(f_{j}^{(1)}, f_{j}^{(2)}, ..., f_{j}^{(h)}\right)$ and $r_{i}$ is set of de-fuzzified values\footnote{$d: D \mid d(A_{ki})=(1/4) (a_{ki} + 2\times b_{ki} + c_{ki})$ for triangular set $A_{ki}$} of antecedent fuzzy sets in $R_{i}$.
The final inferred output is calculated as
\begin{equation}
O^{*}_{j} = \sum\limits_{i=1}^{c} W_{i} \times D_{f}\left(B_{i}\right)
\label{eq:final_oj}
\end{equation}
where $D_{f}\left(B_{i}\right)$ is the de-fuzzified value of consequent fuzzy variables $B_{i}$ with $0 \leq W_{i} \leq 1$ and $\sum_{i=1}^{c} W_{i} = 1$. The type for the traffic is assigned on the basis of inferred output.

\section{Dataset}
\label{sec:dataset}

The data set was collected using a real-world testbed with 30+ typical user devices. These devices include smartphones, tablets, smart appliances and personal computing devices etc., running popular operating systems including iOS, Android, Windows, MAC OS, Tizen and webOS. All devices support wireless connectivity with 32 devices supporting Bluetooth as well.

\textbf{Testbed setup:}
The testbed represents a typical SOHO network, where all user devices connected to an AP through wired/wireless medium and the AP is connected to Internet.
Data was collected by connecting all devices to an AP setup running wireless and wired networks, with one interface connected to the Internet. All traffic over wireless and wired interfaces in both LAN and WAN networks was collected. 

\textbf{Scenarios:}
Table~\ref{tbl:scenarios} shows seven different scenarios used for data collection. These scenarios represent \classB and (commonly expected) \classM device activity. Data collection for each scenario was repeated for $n=20$ times to avoid any discrepancies and peculiarities in the data. For each iteration, a set of devices (may vary depending on scenario) was connected to the network and data was collected from both \wifi and Ethernet interfaces, to record all traffic within and across the network, including the traffic among wireless clients. After each iteration, the testbed (including devices) was reset to get a clean-slate for next iteration. Non-overlapping set of devices was used for data collection in similar scenarios, to minimize any redundancy and remove any device specific behaviors from the dataset. Any duplicate data points were removed from the dataset to prevent any bias in the learning algorithm.

\begin{table}[h]
\centering
\caption{Scenarios for data collection, representing network activity types}
\label{tbl:scenarios}
\begin{tabular}{p{0.23\linewidth}p{0.77\linewidth}}
\toprule
\textbf{Scenario}         & \multicolumn{1}{c}{\textbf{Description}}    \\
\midrule
Auth.\ attack (A) & A compromised host makes multiple login attempts to other host(s) 	\\ 
Botnet activity (B)       & A compromised host opens many connections to one or more usually remote destination hosts. 	\\ 
Normal (N)                & Typical, non-malicious, usage pattern \\
Port Sweep (\psweep)      & A compromised host scans all ports on a destination host.      \\
Port Scan (\pscan)       & A compromised host scans a subset of all ports of a target.		\\ 
Spying (S)                & A compromised host tries to send user data to a remote destination.  		\\ 
Worm (W)                  & A compromised host scans the network for access to other hosts and tries to copy malicious content on destination host(s).                              	\\
\bottomrule
\end{tabular}
\end{table}

Due to real-world testbed setting, dataset imbalance issues result in \classB traffic becoming \textit{majority} class and \classM traffic becoming \textit{minority} class. It is because devices rarely exhibit \classM behavior~\cite{MaliciousTraffic3, MaliciousTraffic30}. In order to prevent the imbalanced data problem, data points from \textit{majority} class are undersampled. The experiments showed that under-sampling does not affect the accuracy of prediction because the \textit{majority} class data is correlated and under-sampling does not result in loss of significant traits in the data. Meanwhile, \textit{minority} class data points were over-sampled using SMOTE~\cite{Chawla:2002:SSM:1622407.1622416} to get $7:3$ ratio for \textit{benign}:\textit{malicious} class data points. All six sub-classes in \textit{minority} class contain equal data points.

\section{Evaluation}
\label{sec:evaluation}
 
\begin{figure}[t]
\begin{minipage}{.50\linewidth}
\centering
\subfloat{\label{fig:dataset_mds}\includegraphics[trim={40mm 20mm 35mm 30mm},clip,width=0.7\textwidth]{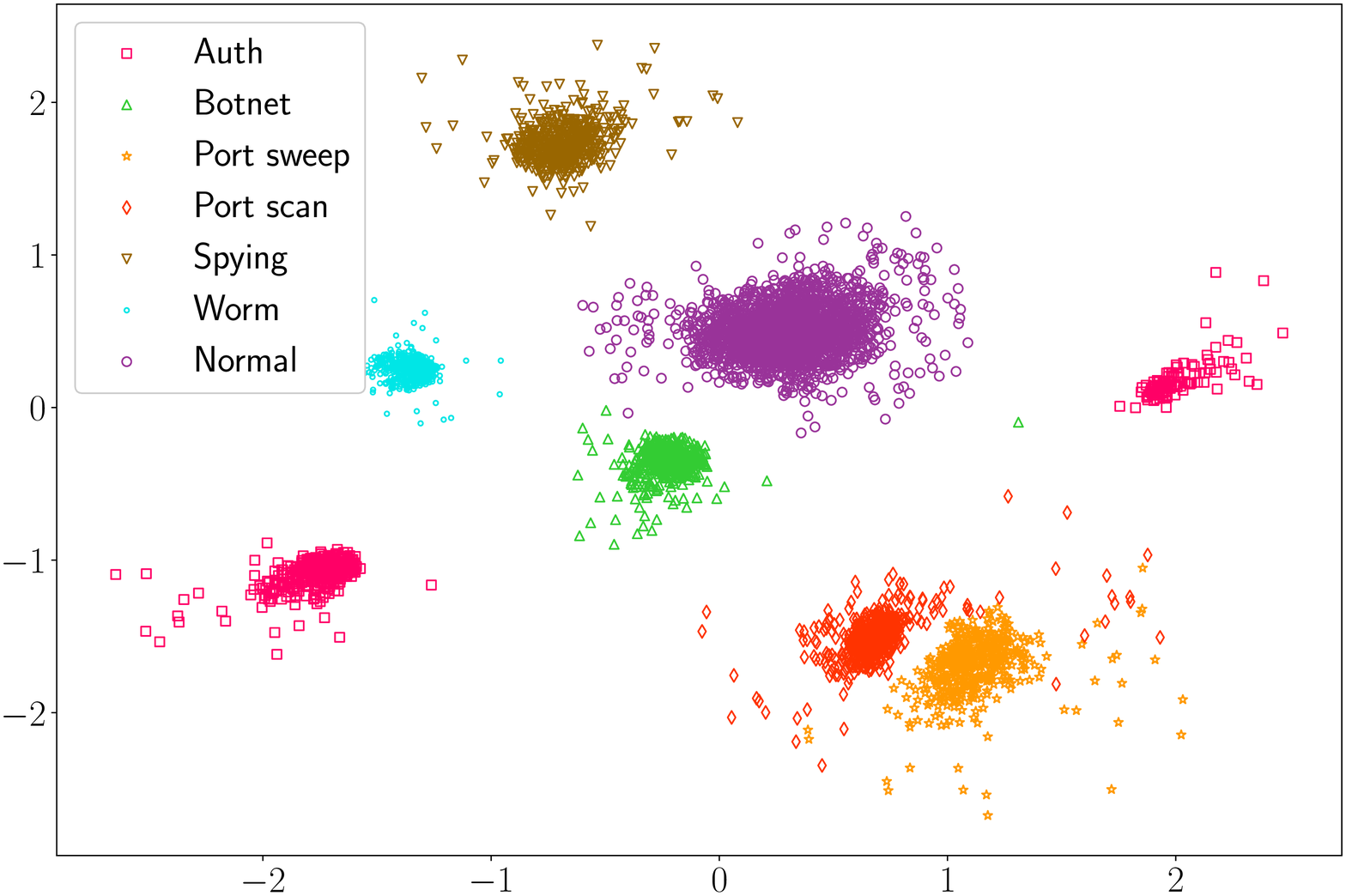}}
\end{minipage}%
\begin{minipage}{.50\linewidth}
\centering
\subfloat{\label{fig:classify_vs_rc}\includegraphics[trim={5mm 10mm 10mm 5mm},clip,width=0.7\textwidth]{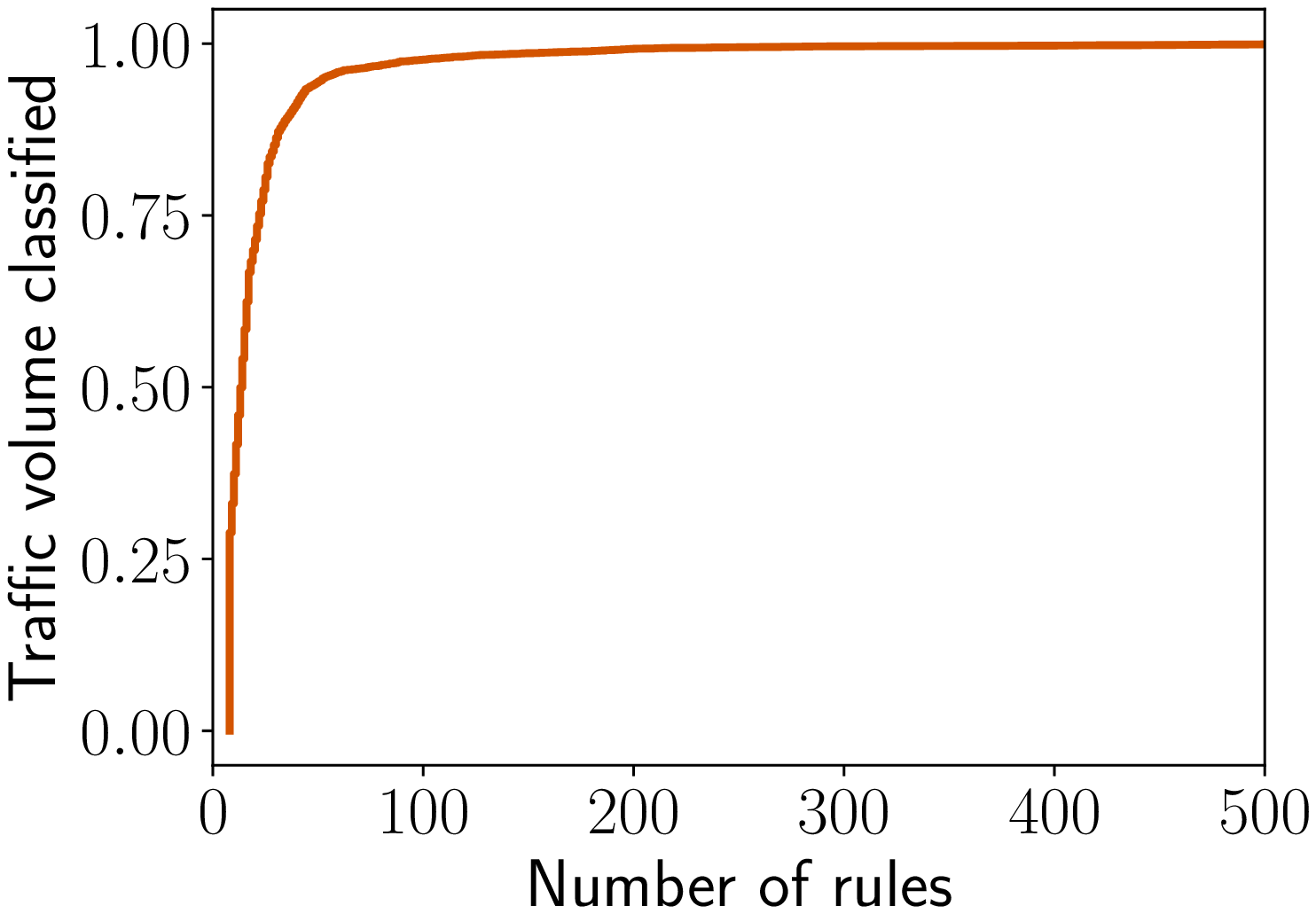}}
\end{minipage} %
\caption{(a): Clusters obtained as a result of applying FCM clustering algorithm. (b): CDF plot for the number of classification rules required to predict traffic class}
\label{fig:clustering}
\end{figure}

Feature extraction, feature analysis, clustering and prediction scheme was implemented with Python using \texttt{dpkt}, \texttt{imbalanced-learn} and \texttt{scikit-learn} libraries.  
After feature reduction, clustering was performed to groups all the data points into clearly differentiable clusters based on self-similarity. Figure~\ref{fig:dataset_mds} shows the clusters obtained by performing FCM clustering on our dataset. The figure was plotted by mapping 22 dimensional feature space to 2 dimensional surface using \textit{multi-dimensional scaling} (MDS)~\cite{REHMAN2018149}. Figure~\ref{fig:dataset_mds} shows that our technique produces clearly differentiable clusters with distinct boundaries. 
Moreover, the figure shows that the clustering algorithm can also easily among distinguish different sub-classes of \classM traffic. 
After clustering, the feature value distributions in each of the clusters are shown in Figure~\ref{fig:dataset_mds}. 

\begin{figure}[t]
\begin{minipage}{.33\linewidth}
\centering
\subfloat{\label{fig:nf_port_sweep}\includegraphics[scale=.22]{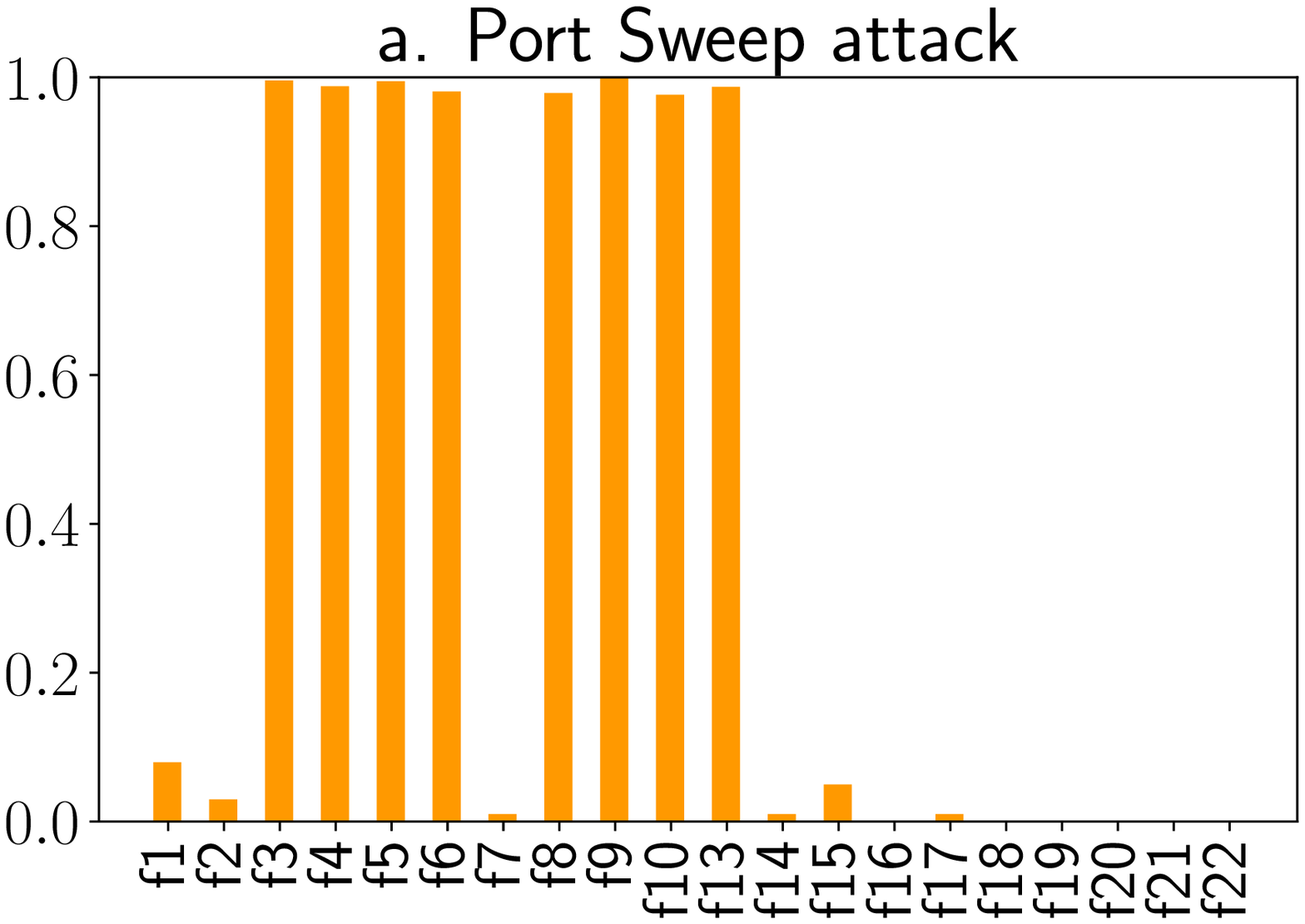}}
\end{minipage}%
\begin{minipage}{.33\linewidth}
\centering
\subfloat{\label{fig:nf_port_scan}\includegraphics[scale=.22]{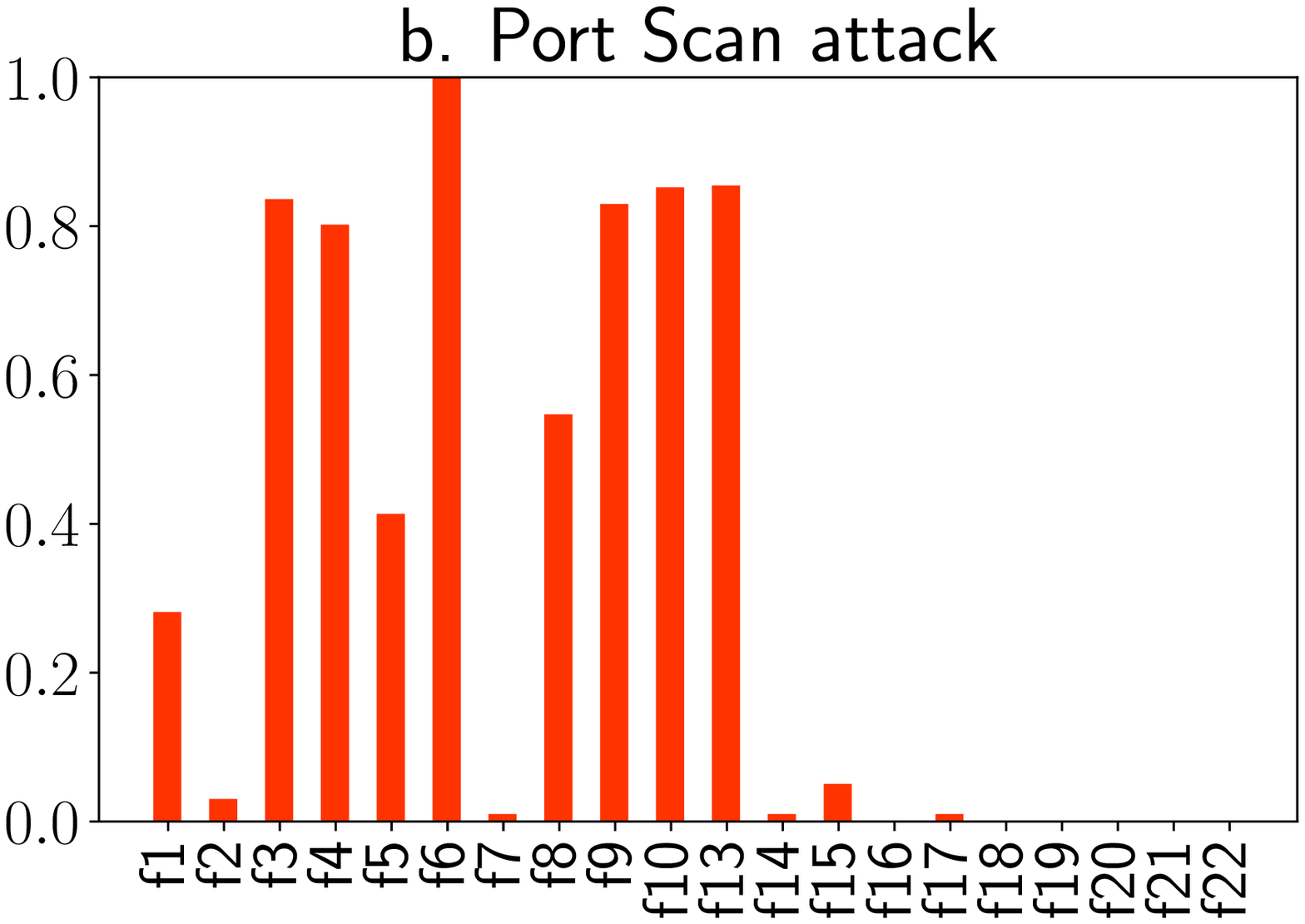}}
\end{minipage} %
\begin{minipage}{.33\linewidth}
\centering
\subfloat{\label{fig:nf_auth}\includegraphics[scale=.22]{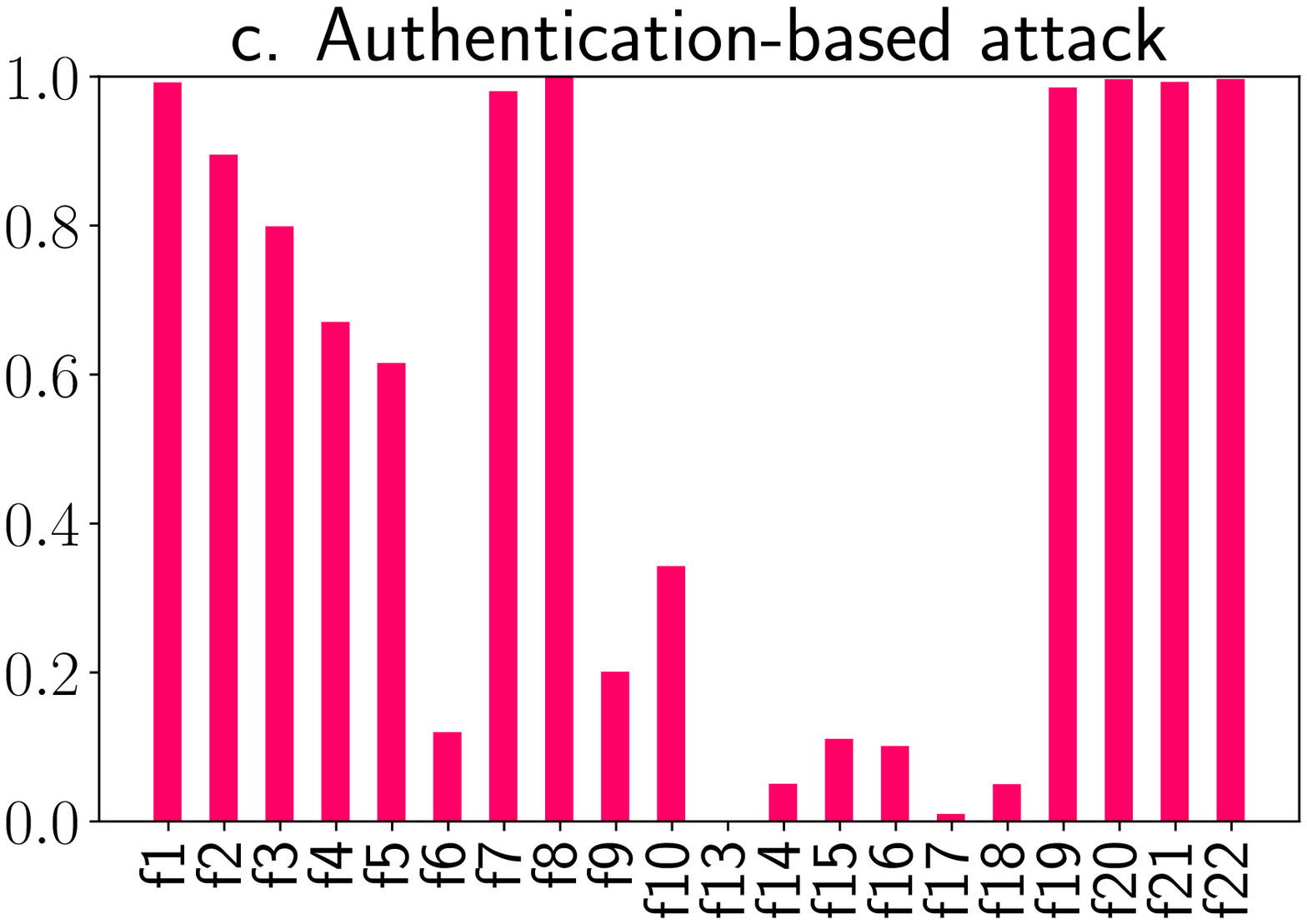}}
\end{minipage}

\begin{minipage}{.33\linewidth}
\centering
\subfloat{\label{fig:nf_worm}\includegraphics[scale=.22]{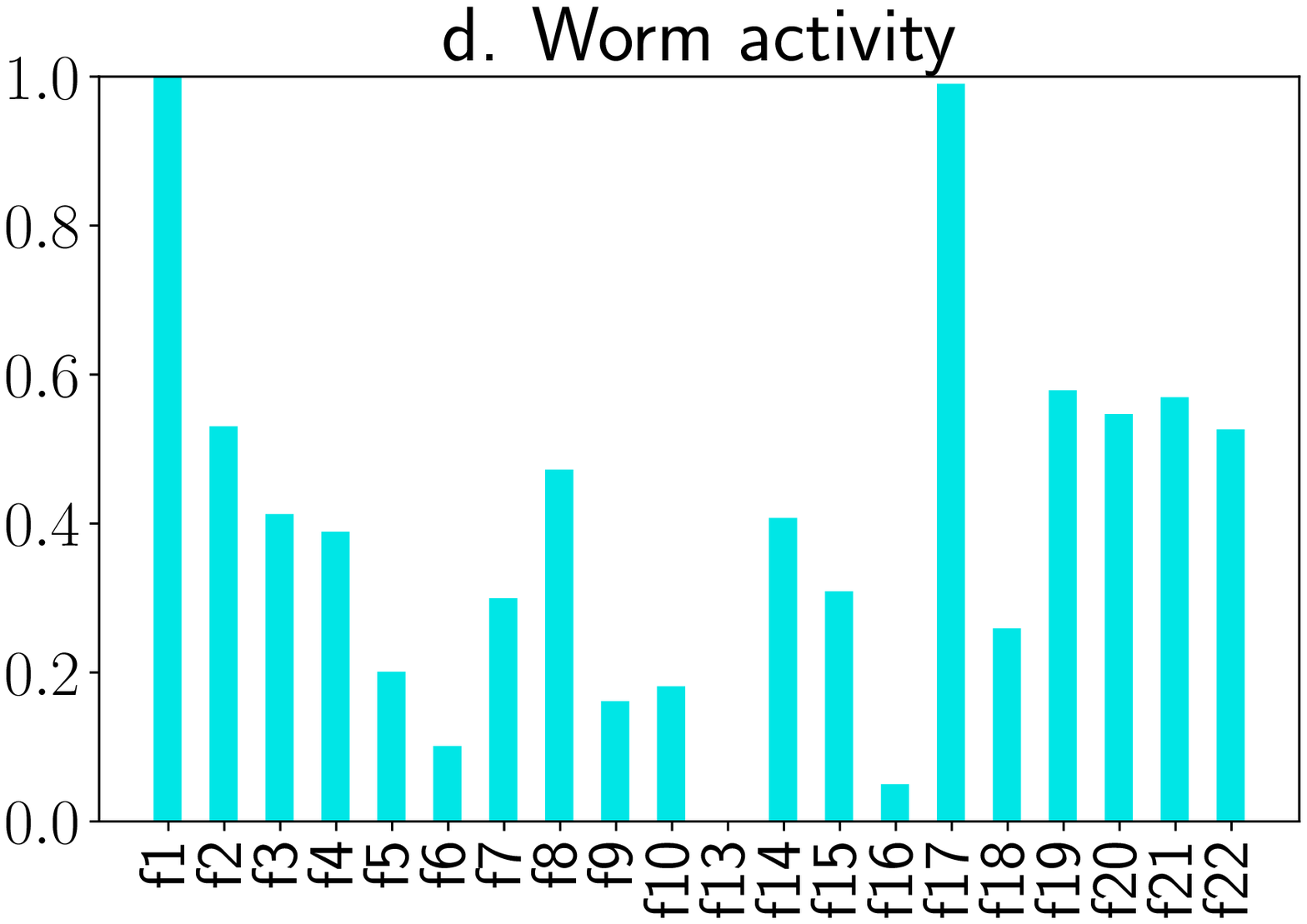}}
\end{minipage}%
\begin{minipage}{.33\linewidth}
\centering
\subfloat{\label{fig:nf_botnet}\includegraphics[scale=.22]{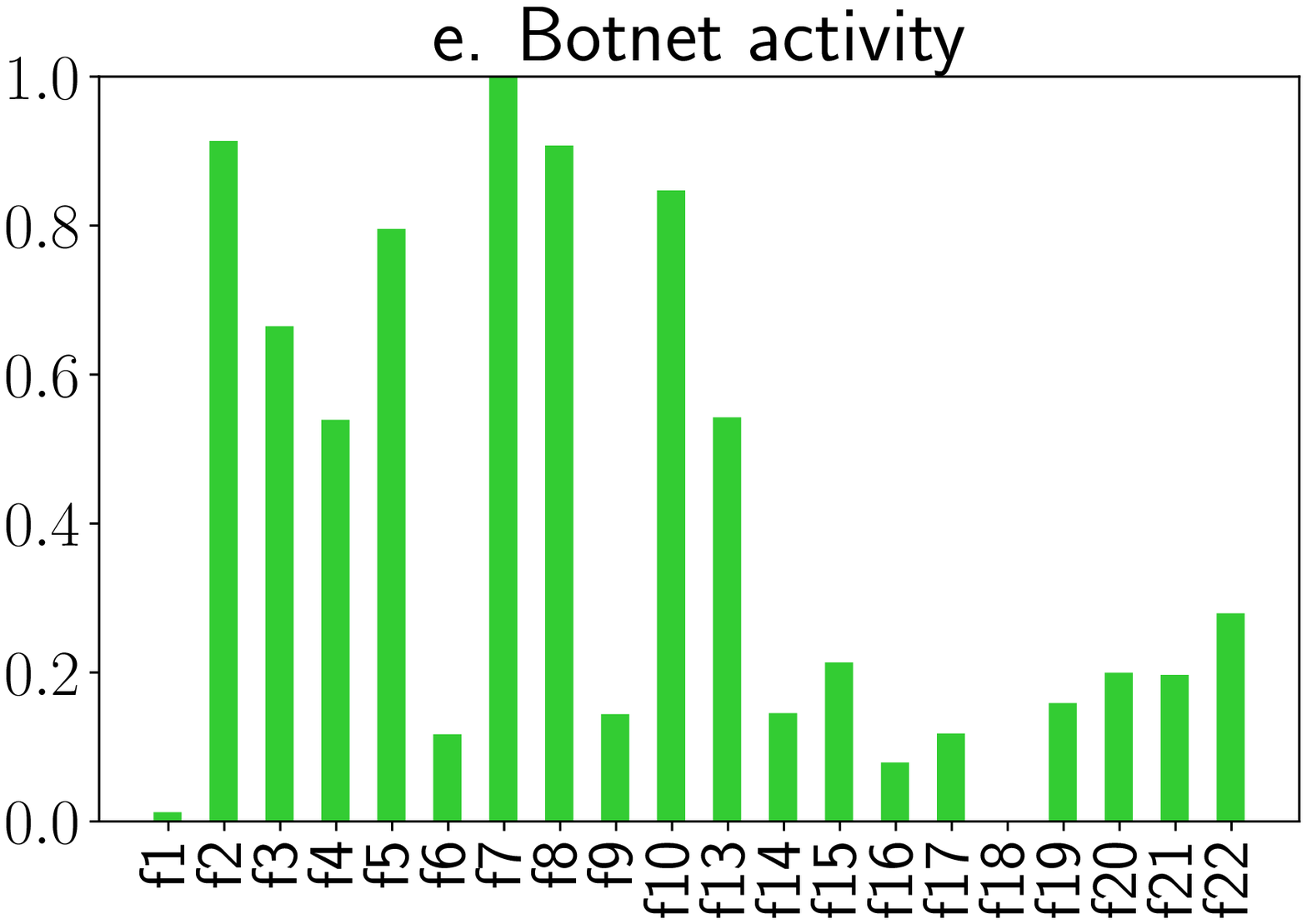}}
\end{minipage}%
\begin{minipage}{.33\linewidth}
\centering
\subfloat{\label{fig:nf_spying}\includegraphics[scale=.22]{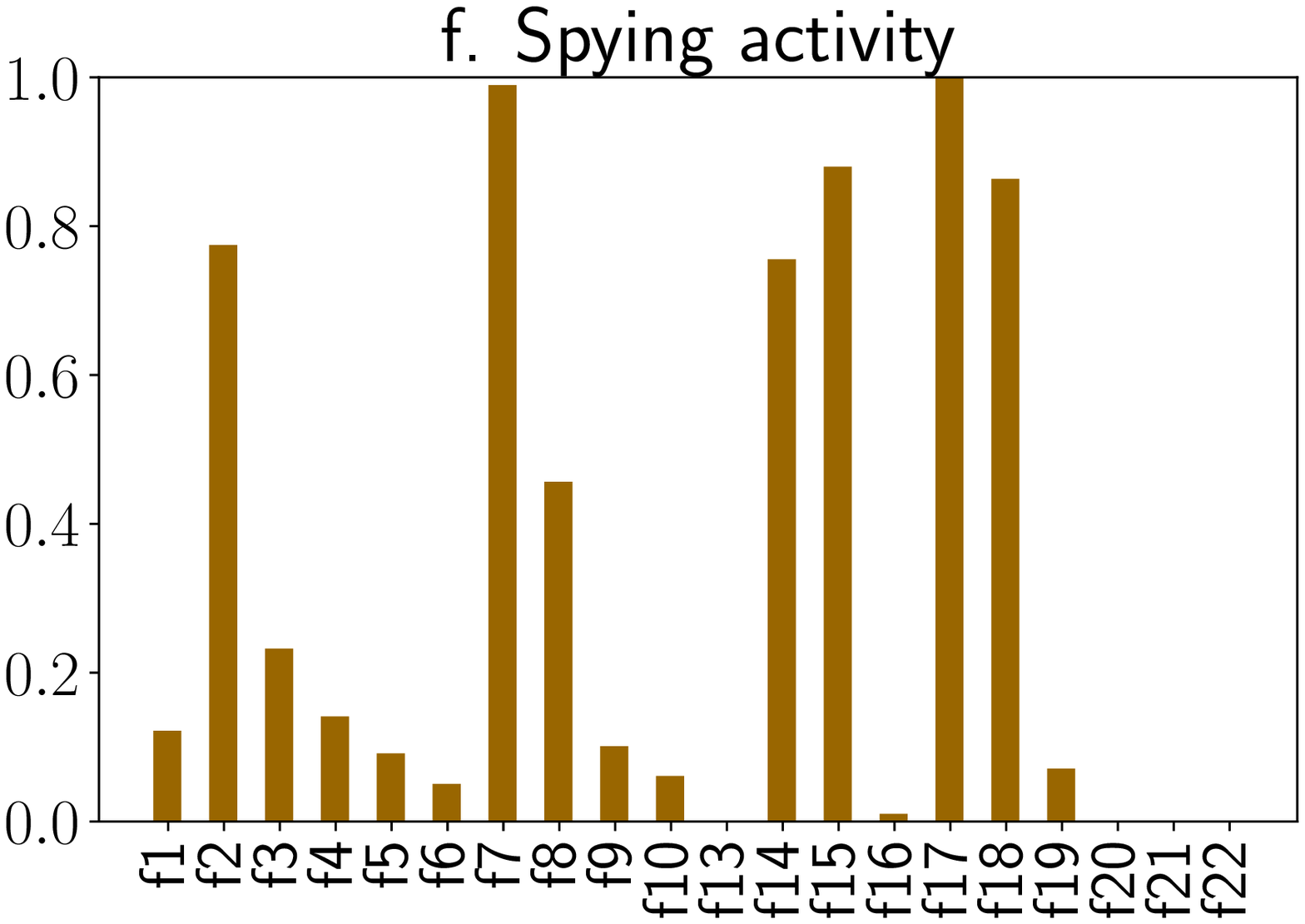}}
\end{minipage}
\caption{Normalized feature averages for distinguishing features representing clusters for each attack scenario.}
\label{fig:nf_features}
\end{figure}

Figure~\ref{fig:nf_features} show clearly distinguishable normalized features scores for each of the clusters. Out of the 39 features extracted from network metadata (see Sect.~\ref{sec:feature_extraction}), the normalized feature scores of 18 features was studied.
Features \texttt{f1-f9} correspond to connection-related data (e.g.\ connection count, unique IPs), \texttt{f10-f13} correspond to flagged packets (e.g.\ urgent, SYN, REJ), \texttt{f14-f18} correspond to data (e.g.\ SRC2DST, DST2SRC), and \texttt{f19-f22} correspond to authentication related features (e.g. SSH connections, login attempts).

Table~\ref{tbl:conf_mat_2class} shows the prediction accuracy for binary-class problem, where \sysname achieves $98.61\%$ accuracy with a \textit{precision} of $0.985$ and $0.99$ score for both \textit{sensitivity} and \textit{specificity}, giving us an overall F1-score of $0.986$. 

\begin{table}[h]
\caption{Confusion matrix for binary-class problem}
\label{tbl:conf_mat_2class}
\centering
\begin{tabular}{l|l|c|c|c}
\multicolumn{2}{c}{}&\multicolumn{2}{c}{\textit{Predicted}}&\\
\cline{3-4}
\multicolumn{2}{c|}{}&\textbf{\classB}&\textbf{\classM}&\multicolumn{1}{c}{\textbf{Total}}\\
\cline{2-4}
\multirow{2}{*}{\textit{Actual}}& \textbf{\classB} & $533$ & $7$ & $540$\\
\cline{2-4}
& \textbf{\classM} & $8$ & $532$ & $540$\\
\cline{2-4}
\multicolumn{1}{c}{} & \multicolumn{1}{c}{\textbf{Total}} & \multicolumn{1}{c}{$541$} & \multicolumn{1}{c}{$539$} & \multicolumn{1}{c}{$1080$}\\
\end{tabular}
\end{table} 

The decision to preserve information in outliers (tail of distribution) helped in clearly differentiating between otherwise overlapping classes, resulting in good prediction accuracy.
By removing the features containing redundant information, FCM algorithm was able to generate clusters with distinct boundaries, resulting in low false positives and false negatives. 
The choice of FCM algorithm was also helpful in improving accuracy because it allowed us to use the weighted association with neighbouring clusters to predict class labels for new traffic flows. 

Table~\ref{tbl:conf_mat_6class} shows the prediction accuracy for detecting individual attack types. It shows that, on average, \sysname achieved $98\%$ accuracy with $0.94$ F1-score in multi-class prediction. 

\begin{table}[h]
\centering
\caption{(a): Confusion matrix for the \classM activity types. A=Actual, P=Predicted; (b): Prediction performance for subclasses of \classM traffic}
\subfloat[]{\label{tbl:conf_mat_6class}
\begin{tabular}{lllllll}
\multicolumn{1}{l|}{\backslashbox{A}{P}} & A           & B           & PS          & Ps          & S           & W           \\ \hline
\multicolumn{1}{l|}{A}   & \textbf{86} & 0           & 0           & 0           & 0           & 3            \\
\multicolumn{1}{l|}{B}   & 0           & \textbf{83} & 0           & 2           & 2           & 0            \\
\multicolumn{1}{l|}{PS}  & 0           & 0           & \textbf{81} & 9	       & 0           & 0            \\
\multicolumn{1}{l|}{Ps}  & 3           & 0           & 3          & \textbf{84}  & 0           & 0            \\
\multicolumn{1}{l|}{S}   & 0           & 0           & 0           & 0           & \textbf{86} & 1            \\
\multicolumn{1}{l|}{W}   & 3           & 2           & 0           & 0           & 0           & \textbf{84} 
\end{tabular}}~~~~

\subfloat[]{\label{tbl:classifier_measures}
\begin{tabular}{l c c c c c c c c}&\\
\multicolumn{1}{l|}{Measure/}	& A     & B		& PS    & Ps	 & S     & W     & Mean  \\ \hline
\multicolumn{1}{l|}{Accuracy}          & 0.98  & 0.98	& 0.98  & 0.96  & 0.98  & 0.98  & 0.98  \\
\multicolumn{1}{l|}{Precision}         & 0.93  & 0.95	& 0.96  & 0.87  & 0.95  & 0.95  & 0.94 \\
\multicolumn{1}{l|}{Specificity} 	    & 0.96  & 0.92	& 0.90  & 0.93  & 0.96  & 0.93  & 0.94 \\
\multicolumn{1}{l|}{Sensitivity} 	    & 0.99  & 0.99	& 0.99  & 0.97  & 0.99  & 0.99  & 0.99 \\
\multicolumn{1}{l|}{F1-score}          & 0.95  & 0.94	& 0.93  & 0.90  & 0.95  & 0.94  & 0.94 \\
&\\
\end{tabular}}
\end{table}

Although \sysname was able to predict \psweep and \pscan with high accuracy, the highest number of inaccurate predictions have been made for these two classes. Figure~\ref{fig:nf_port_sweep} and Figure~\ref{fig:nf_port_scan} show that this behaviour is due to overlapping the feature scores of these classes e.g., the instances of \pscan attack over a large range of ports will result in so many connections that it is predicted as \psweep attack. Similarly, a \psweep attack over a limited range of ports may be predicted as \pscan attack.

\section{Comparison with existing approaches}
\label{sec:stoa}
A number of researchers have used machine learning (ML) algorithms to detect malicious traffic~\cite{4738466, 7346821}. Their proposals use data mining~\cite{Jeyakumar:2016:DDD:2875475.2875490}, supervised ML ~\cite{6510197}, and unsupervised ML techniques~\cite{nids-fuzzy-2011, Bohara:2016:IDE:2898375.2898400} to build network intrusion detection systems (NIDS).
Bekerman et al.~\cite{7346821} used 942 features to identify malware by analysing network traffic. 
Strayer et al.~\cite{Strayer2008} and Lu et al.~\cite{Lu:2009:ADB:1533057.1533062} studied network behaviour and application classification to identify bots in networks.
BotMiner~\cite{Gu:2008:BCA:1496711.1496721} used clustering technique for detecting botnets independent of underlying command-and-control protocol and strategy.
Bohara et al.~\cite{Bohara:2016:IDE:2898375.2898400} used unsupervised learning to predict class labels for unlabelled network.

IoT Sentinel~\cite{7980167} uses device type information to limit the network access of vulnerable devices whereas PorfilIoT~\cite{Meidan:2017:PML:3019612.3019878} uses a multi-stage classification technique for differentiating IoT from non-IoT devices and then finding the actual type of IoT device.
IoT Sentinel and ProfilIoT rely on fingerprints generated from devices' network activity to train classification models, and thus fail to detect impersonation attacks. 
Roux et al.~\cite{roux:hal-01561710} propose the use of RSSI using radio probes to detect an attacker trying to hijack user devices. Cheng et. al~\cite{7981520} propose running time patching of access points to block malicious traffic flows in the network.
Barrera et al.~\cite{idiot-arxiv} proposed a security policy enforcement framework for restricting IoT devices communication to necessary interactions.

\begin{table}[th]
\centering
\caption{Qualitative comparison of anomaly detection techniques}
\label{tbl:qualitiative_comparison}
\begin{tabular}{l ccc}
\toprule
\textbf{System} 									& Feature count	& Learning algorithm		&  Functionality  \\
\midrule
Strayer et al.~\cite{Strayer2008}         			& 16	& supervised   &  Botnet detection  \\
IoT Sentinel~\cite{7980167}							& 23	& supervised	&    Device identification   \\
Beckerman et al.~\cite{7346821} 					& 972	& supervised 	&  Malware detection    \\
Yi et al.~\cite{8322605}							& 5 	& supervised 	& Anomaly detection\\
Median et al.~\cite{DBLP:journals/corr/abs-1709-04647}	& 274 	& supervised	& Device identification \\
\sysname   											& 18-39	& semi-supervised   & Anomaly detection\\
\bottomrule
\end{tabular}
\end{table}

Meidan et al.~\cite{Meidan:2017:PML:3019612.3019878} used machine learning approach for detecting device types for 17 IoT devices with $99.4\%$ accuracy.
Ran et al.~\cite{8322736} proposed a self-adaptive technique for traffic classification based on semi-supervised machine learning, which dynamically choose optimal system parameters to achieve high accuracy.
Yi et al.~\cite{8322605} proposed an algorithm using decision trees (DT) and co-training to detect abnormal/botnet traffic generated by a webcam. 

Table~\ref{tbl:qualitiative_comparison} presents a qualitative comparison of \sysname with current state of the art in traffic classification and IoT security research. \sysname is considered \textit{semi-supervised} only because the labels assigned to the clusters are manually verified.

\section{Discussion}
\label{sec:discussion}

The evaluation of \sysname shows that our approach allows us to successfully accurately predict the various type of traffic, discussed in our threat model.

The ability to use unlabelled data can be useful in improving the traffic classification schemes for a number of reasons. It will save the effort of labelling all traffic data used for model training and makes the system more flexible to adapt. 

This work mainly use the data extracted from network and device logs. Therefore, our technique can be used in various network settings irrespective of what devices are connected to the network. Our model allows us to extend the classification to identify more types of \classM traffic seen in the network e.g. crypto-jacking attacks, which usually exhibit traffic patterns as seen in spying attacks.
\sysname can also be extended further to classify the sub-types of normal user traffic. This information could then be used for on-demand bandwidth provisioning and dynamic traffic management based on the traffic patterns. 

A possible limitation of \sysname is that it can only identify a device's malicious activity if it communicates over the network. That is, \sysname cannot tell if an attacker physically accessed a device e.g. smart door-bell, and extracted information by directly connecting to it over physical, serial connection. 
However, \sysname will be able to identify the (misbehaving) tampered device as soon as it connects to the network, and prevents it from executing any attacks against local or remote destinations.

\sysname has been evaluated using devices with both wireless and wired network connection. However, its performance is not analysed for lower power communication protocols such as Zigbee, Z-Wave, Bluetooth LE etc.
The process of verifying labels assigned to clusters can be automated, by cross-referencing the information from other sources. We expect the future research to explore new set of features to extend the types of attacks this approach can classify. 

Finally, software updates in devices may change their network behaviour, which can initially be detected as malicious. However, \sysname can adapt to this new network behaviour quickly, to stop prevent any false-positives. This behaviour was intentional, as it can be used to detect firmware versions of devices connected to the network.

\section{Conclusion}
\label{sec:conclusion}

This paper present a lightweight semi-supervised learning based technique, \sysname, for identifying \classB and several types of \classM traffic in edge networks.
This paper introduces a threat model based on the most common attacks in IoT landscape and a real-world testbed setup for collecting network data and device level logs. 
Our proposed pipeline for feature extraction, analysis, and reduction identifies the set of features that yield most value to the IoT device activity detection.
Evaluation results for \sysname show that using clustering and FIS, various types of \classM and \classB traffic can be predicted with high accuracy in real-time. 

In specific, \sysname is able to predict traffic class in 250\,ms, without requiring specialized hardware. 
\sysname can be extended to identify more attack types, other than the ones considered in this paper. The technique can also be re-purposed for detecting devices, firmware versions and improving network bandwidth management, traffic routing problem based on the traffic patterns observed in the networks. 

\section*{Acknowledgements}
The work was supported in part by the Business Finland PraNA research project.

%
%
%
\bibliographystyle{splncs04}
\bibliography{iotguard}

\begin{thebibliography}{10}
\providecommand{\url}[1]{\texttt{#1}}
\providecommand{\urlprefix}{URL }
\providecommand{\doi}[1]{https://doi.org/#1}

\bibitem{KDDCup99Dataset}
Kdd cup 1999 data.
  \url{http://kdd.ics.uci.edu/databases/kddcup99/kddcup99.html}, [Accessed:
  2016-07-18]

\bibitem{Senrio:SingleFlaw}
Senrio. 400,000 publicly available iot devices vulnerable to single flaw.
  \url{https://bit.ly/2Ieghvu}, [Accessed: 2017-05-05]

\bibitem{Agrawal:1994:FAM:645920.672836}
Agrawal, R., Srikant, R.: Fast algorithms for mining association rules in large
  databases. In: Proceedings of the 20th International Conference on Very Large
  Data Bases. pp. 487--499. VLDB '94 (1994)

\bibitem{6510197}
Akbar, et~al.: Improving network security using machine learning techniques.
  In: 2012 IEEE International Conference on Computational Intelligence and
  Computing Research. pp.~1--5 (Dec 2012)

\bibitem{4426662}
Aranganayagi, S., Thangavel, K.: Clustering categorical data using silhouette
  coefficient as a relocating measure. In: International Conference on
  Computational Intelligence and Multimedia Applications (ICCIMA 2007). vol.~2,
  pp. 13--17 (2007)

\bibitem{idiot-arxiv}
Barrera, D., Molloy, I., Huang, H.: {IDIoT}: Securing the internet of things
  like it's 1994. CoRR  \textbf{abs/1712.03623} (2017)

\bibitem{7346821}
Bekerman, et~al.: Unknown malware detection using network traffic
  classification. In: 2015 IEEE Conference on Communications and Network
  Security (CNS). pp. 134--142 (Sept 2015)

\bibitem{Bohara:2016:IDE:2898375.2898400}
Bohara, A., Thakore, U., Sanders, W.H.: Intrusion detection in enterprise
  systems by combining and clustering diverse monitor data. In: Proceedings of
  the Symposium and Bootcamp on the Science of Security. pp. 7--16. HotSos '16
  (2016)

\bibitem{Chawla:2002:SSM:1622407.1622416}
Chawla, N.V., Bowyer, K.W., Hall, L.O., Kegelmeyer, W.P.: Smote: Synthetic
  minority over-sampling technique. J. Artif. Int. Res.  \textbf{16}(1),
  321--357 (Jun 2002)

\bibitem{7981520}
Cheng, et~al.: Traffic-aware patching for cyber security in mobile iot. IEEE
  Communications Magazine  \textbf{55}(7),  29--35 (2017)

\bibitem{Gu:2008:BCA:1496711.1496721}
Gu, G., Perdisci, R., Zhang, J., Lee, W.: Botminer: Clustering analysis of
  network traffic for protocol- and structure-independent botnet detection. In:
  Proceedings of the 17th Conference on Security Symposium. pp. 139--154. SS'08
  (2008)

\bibitem{Jeyakumar:2016:DDD:2875475.2875490}
Jeyakumar, V., Madani, O., ParandehGheibi, A., Yadav, N.: Data driven data
  center network security. In: Proceedings of the 2016 ACM on International
  Workshop on Security And Privacy Analytics. pp. 48--48. IWSPA '16 (2016)

\bibitem{roux:hal-01561710}
Jonathan, et~al.: {Toward an Intrusion Detection Approach for IoT based on
  Radio Communications Profiling}. In: {13th European Dependable Computing
  Conference}. p.~4p. Geneva, Switzerland (Sep 2017)

\bibitem{Lu:2009:ADB:1533057.1533062}
Lu, et~al.: Automatic discovery of botnet communities on large-scale
  communication networks. In: Proceedings of the 4th International Symposium on
  Information, Computer, and Communications Security. pp. 1--10. ASIACCS '09
  (2009)

\bibitem{MaliciousTraffic30}
Martindale, J.: Nearly 30 percent of all web traffic is sent by malicious bots.
  \url{https://www.digitaltrends.com/web/bad-bots-intrnet/}, [Accessed:
  2018-04-06]

\bibitem{MaliciousTraffic3}
McMillan, R.: Up to three percent of internet traffic is malicious, researcher
  says.
  \url{https://www.csoonline.com/article/2122506/data-protection/up-to-three-percent-of-internet-traffic-is-malicious--researcher-says.html},
  [Accessed: 2018-04-06]

\bibitem{DBLP:journals/corr/abs-1709-04647}
Meidan, et~al.: Detection of unauthorized iot devices using machine learning
  techniques. CoRR  \textbf{abs/1709.04647} (2017),
  \url{http://arxiv.org/abs/1709.04647}

\bibitem{Meidan:2017:PML:3019612.3019878}
Meidan, Y., Bohadana, M., Shabtai, A., Guarnizo, J.D., Ochoa, M., Tippenhauer,
  N.O., Elovici, Y.: Profiliot: A machine learning approach for iot device
  identification based on network traffic analysis. In: Proceedings of the
  Symposium on Applied Computing. pp. 506--509. SAC '17 (2017)

\bibitem{7980167}
Miettinen, et~al.: Iot sentinel: Automated device-type identification for
  security enforcement in iot. In: 2017 IEEE 37th International Conference on
  Distributed Computing Systems (ICDCS). pp. 2177--2184 (June 2017)

\bibitem{NARVEKAR2015101}
Narvekar, M., Syed, S.F.: An optimized algorithm for association rule mining
  using fp tree. Procedia Computer Science  \textbf{45}(Supplement C),  101 --
  110 (2015),
  \url{http://www.sciencedirect.com/science/article/pii/S1877050915003336},
  international Conference on Advanced Computing Technologies and Applications

\bibitem{4738466}
Nguyen, T.T.T., Armitage, G.: A survey of techniques for internet traffic
  classification using machine learning. IEEE Communications Surveys Tutorials
  \textbf{10}(4),  56--76 (Fourth 2008)

\bibitem{IoTdeviceForecast1}
Nordum, A.: Popular internet of things forecast of 50 billion devices by 2020
  is outdated. \url{https://bit.ly/2K2Tk3Z}, [Accessed: 2017-05-07]

\bibitem{6975580}
Patton, et~al.: Uninvited connections: A study of vulnerable devices on the
  internet of things (iot). In: 2014 IEEE Joint Intelligence and Security
  Informatics Conference. pp. 232--235 (Sept 2014)

\bibitem{DLink:SingleFlaw}
Pauli, D.: 414,949 d-link cameras, iot devices can be hijacked over the net.
  \url{https://www.theregister.co.uk/2016/07/08/414949_dlink_cameras_iot_devices_can_be_hijacked_over_the_net/},
  [Accessed: 2017-05-07]

\bibitem{8322736}
Ran, J., Kong, X., Lin, G., Yuan, D., Hu, H.: A self-adaptive network traffic
  classification system with unknown flow detection. In: 2017 3rd IEEE
  International Conference on Computer and Communications (ICCC). pp.
  1215--1220 (Dec 2017)

\bibitem{REHMAN2018149}
ur~Rehman, Z., Idris, A., Khan, A.: Multi-dimensional scaling based grouping of
  known complexes and intelligent protein complex detection. Computational
  Biology and Chemistry  \textbf{74},  149 -- 156 (2018).
  \doi{https://doi.org/10.1016/j.compbiolchem.2018.03.023}

\bibitem{5370013}
Shanmugam, B., Idris, N.B.: Improved intrusion detection system using fuzzy
  logic for detecting anamoly and misuse type of attacks. In: 2009
  International Conference of Soft Computing and Pattern Recognition. pp.
  212--217 (Dec 2009)

\bibitem{nids-fuzzy-2011}
Shanmugavadivu, R., Nagarajan, N.: Network intrusion detection system using
  fuzzy logic. Indian Journal of Computer Science and Engineering (IJCSE)
  \textbf{2}(1),  101--111 (2001)

\bibitem{Strayer2008}
Strayer, et~al.: Botnet Detection Based on Network Behavior, pp. 1--24. Boston,
  MA (2008)

\bibitem{TRAUWAERT1988217}
Trauwaert, E.: On the meaning of dunn's partition coefficient for fuzzy
  clusters. Fuzzy Sets and Systems  \textbf{25}(2),  217 -- 242 (1988)

\bibitem{8322605}
Yi, L., Shi, Y.: Research on abnormal traffic classification of web camera
  based on supervised learning and semi-supervised learning. In: 2017 3rd IEEE
  International Conference on Computer and Communications (ICCC). pp. 547--551
  (2017)

\bibitem{Zhou2014}
Zhou, et~al.: Fuzziness parameter selection in fuzzy c-means: The perspective
  of cluster validation. Science China Information Sciences  \textbf{57}(11),
  ~1--8 (2014)

\end{thebibliography}

\end{document}